\documentclass[twoside,reqno]{HERON}
\usepackage{epsfig,cite,colordvi}
\usepackage{graphicx}
\usepackage{url}
\usepackage{amssymb,amsmath,amscd,epsf}
\usepackage{times}
\usepackage{makeidx}
\begin{document}

\title{Manifestations of SU(3) symmetry in heavy deformed nuclei
}

\runningheads{SU(3) symmetry in heavy deformed nuclei
}{D. Bonatsos, A. Martinou, I.E. Assimakis, S. Sarantopoulou, S. Peroulis, \it{et al.}}

\begin{start}

\author{Dennis Bonatsos}{1}, \coauthor{Andriana Martinou}{1}, \coauthor{I.E. Assimakis}{1}, 
\coauthor{S. Sarantopoulou}{1}, \coauthor{S. Peroulis}{2}, \coauthor{N. Minkov}{3} 

\index{Bonatsos, D.}
\index{Martinou, A.}
\index{Assimakis, I.E.}
\index{Sarantopoulou, S.}
\index{Peroulis, S.}
\index{Minkov, N.}

\address{Institute of Nuclear and Particle Physics, National Centre for Scientific Research 
``Demokritos'', GR-15310 Aghia Paraskevi, Attiki, Greece}{1}

\address{Department of Physics, National and Kapodistrian University of Athens, Zografou Campus, GR-15784 Athens, Greece}{2}

\address{Institute of Nuclear Research and Nuclear Energy, Bulgarian Academy of Sciences, 72 Tzarigrad Road, 1784 Sofia, Bulgaria}{3}

\begin{Abstract}
The rapid increase of computational power over the last several years has allowed detailed microscopic investigations of the structure of many nuclei in terms of Relativistic Mean Field theories as well as in the framework of the no-core Shell Model. In heavy deformed nuclei, in which microscopic calculations remain a challenge, algebraic models based on the SU(3) symmetry offer specific predictions, parameter-independent in several cases, directly comparable to experimental data. Two different approximate models for heavy deformed nuclei based on the SU(3) symmetry, the pseudo-SU(3) and the proxy-SU(3) schemes will be discussed and the compatibility between their predictions for the nuclear deformation parameters will be shown. In particular, the dominance of prolate over oblate shapes in the ground states of even-even nuclei and the prolate to oblate shape phase transition occurring in heavy rare earths will be considered.
\end{Abstract}
\end{start}

Since the path breaking work of Elliott \cite{Elliott1,Elliott2,Elliott3} in the sd nuclear shell, the SU(3) symmetry originating from the three-dimensional harmonic oscillator \cite{Wybourne} has been widely used in nuclear physics \cite{IA,RW}. However, the spin-orbit interaction is known \cite{Nilsson1,Nilsson2} to break the SU(3)  symmetry in the nuclear shells beyond the sd shell, by lowering from each nuclear shell the orbital with the highest angular momentum into the nuclear shell below and bringing into the shell the analogue orbital from the nuclear shell above, which possesses one additional harmonic oscillator quantum and therefore has opposite parity, 
called the abnormal parity orbital \cite{DW1}. Restoration of the SU(3) symmetry in higher nuclear shells has been achieved within the pseudo-SU(3) scheme \cite{pseudo1,pseudo2}, in which the remaining orbitals
in a shell, called the normal parity orbitals \cite{DW1}, are mapped through a unitary transformation 
\cite{Quesne} onto the full nuclear shell possessing one harmonic oscillator quantum less.     

Within each nuclear shell in the pseudo-SU(3) scheme the normal parity protons (neutrons)  and the abnornal parity protons (neutrons) are counted separately in the proton (neutron) valence shell, since the former obey the pseudo-SU(3) symmetry, while the latter do not, since they belong to a single $j$-shell, where $j$ is the total angular momentum \cite{DW1}. The symmetry of the normal parity protons (neutrons) is characterized by the appropriate irreducible representation (irrep) of SU(3), while the abnormal parity protons (neutrons) are considered as spectators.
In Elliott's notation, the SU(3) irreps are characterized by $(\lambda, \mu)$, where $\lambda$ and $\mu$ 
are the Elliott quantum numbers \cite{Elliott1,Elliott2,Elliott3}. If $(\lambda_p,\mu_p)$ is the irrep 
characterizing the normal parity valence protons and  $(\lambda_n,\mu_n)$ is the irrep 
characterizing the normal parity valence neutrons, then the nucleus is characterized by the irrep 
$(\lambda_p+\lambda_n, \mu_p +\mu_n)$ \cite{DW1}. 

For each nucleus one can determine the distribution of protons (neutrons) in normal and abnormal parity levels by looking at the relevant deformation in the appropriate Nilsson diagram \cite{Nilsson1,Nilsson2}.
In what follows we are going to use the deformations predicted by the D1S Gogny interaction \cite{Gogny}, 
which are in very good agreement with the experimental values \cite{Pritychenko}, where they exist. 

An important question regards the irrep which should be used. Since the quadrupole-quadrupole interaction plays a leading role in deformed nuclei \cite{Elliott1,Elliott2,Elliott3}, the irrep in which this interaction is maximized has been used over the years, called the most leading irrep\cite{DW1,pseudo1,pseudo2}. Since the quadrupole-quadrupole interaction is simply related to the second order Casimir operator of SU(3), $C_2^{SU(3)}$, through \cite{DW1}
\begin{equation}
Q\cdot Q = 4 C_2^{SU(3)} - 3 L\cdot L, 
\end{equation}
where $Q$ is the quadrupole operator and $L$ the angular momentum operator, 
the irrep 
with the highest eigenvalue of $C_2^{SU(3)}$, given by \cite{DW1}
 \begin{equation}\label{C2} 
 C_2^{SU(3)}(\lambda,\mu)= \lambda^2+\lambda \mu + \mu^2+ 3\lambda +3 \mu, 
\end{equation}
 has been used as the most leading irrep.
 

\begin{table}[htb]

\caption{Distribution of valence protons and valence neutrons into normal and abnormal parity orbitals in the rare earth region, as obtained from the standard Nilsson diagrams \cite{Lederer}, using for each nucleus the deformation parameter obtained from Ref. \cite{Gogny}.  
}

\rotatebox{90}{

\begin{tabular}{ r c c c c c c c c c c c c c  }

\hline
  & Xe & Ba & Ce & Nd & Sm & Gd & Dy & Er & Yb & Hf & W & Os & Pt \\
Z$_{val}$ & 4  & 6  &  8 & 10 & 12 & 14 & 16 & 18 & 20 & 22 & 24 & 26 & 28 \\
   &4+0& 6+0 & 6+2&6+4 &8+4 &10+4&10+6&10+8&12+8&14+8&16+8&16+10&16+12\\

\hline
 N$_{val}$&     &     &     &     &     &     &     &     &     &     &     &     &     \\
 2&  2+0&  2+0&  2+0&  2+0&  2+0&  2+0&  2+0&  2+0&  2+0&  2+0&  2+0&  2+0&  2+0\\
 4&  4+0&  4+0&  4+0&  4+0&  4+0&  4+0&  4+0&  4+0&  4+0&  4+0&  4+0&  4+0&  4+0\\
 6&  6+0&  6+0&  6+0&  6+0&  6+0&  6+0&  6+0&  6+0&  6+0&  6+0&  6+0&  6+0&  6+0\\
 8&  8+0&  6+2&  6+2&  6+2&  6+2&  6+2&  6+2&  6+2&  6+2&  8+0&  8+0&  8+0&  8+0\\
10&  8+2&  6+4&  6+4&  6+4&  6+4&  6+4&  6+4&  6+4&  6+4&  8+2&  8+2&  8+2& 10+0\\
12&  8+4&  8+4&  8+4&  8+4&  8+4&  8+4&  8+4&  8+4&  8+4& 10+2&  8+4&  8+4& 10+2\\
14& 10+4&  8+6&  8+6&  8+6&  8+6&  8+6&  8+6&  8+6&  8+6& 10+4& 10+4&  8+6& 10+4\\
16& 10+6& 10+6& 10+6& 10+6& 10+6& 10+6& 10+6& 10+6& 10+6& 12+4& 10+6& 10+6& 10+6\\
18& 12+6& 12+6& 12+6& 12+6& 12+6& 12+6& 12+6& 12+6& 12+6& 12+6& 12+6& 12+6& 12+6\\
20& 12+8& 12+8& 12+8& 12+8& 12+8& 12+8& 12+8& 12+8& 14+6& 14+6& 12+8& 12+8& 12+8\\
22& 14+8& 14+8& 14+8& 14+8& 14+8& 14+8& 14+8& 14+8& 14+8& 16+6& 14+8& 14+8& 14+8\\
24& 16+8& 16+8& 16+8& 16+8& 16+8& 16+8& 16+8& 16+8& 16+8& 16+8& 16+8&14+10& 16+8\\
26&16+10&16+10&16+10&16+10&16+10&16+10&16+10&16+10& 18+8&16+10&16+10&16+10&16+10\\
28&18+10&18+10&18+10&18+10&18+10&18+10&18+10&18+10&18+10&18+10&18+10&18+10&18+10\\
30&20+10&20+10&20+10&20+10&20+10&20+10&20+10&20+10&20+10&20+10&20+10&20+10&20+10\\
32&22+10&22+10&22+10&22+10&22+10&22+10&22+10&22+10&22+10&22+10&22+10&22+10&22+10\\
34&24+10&24+10&24+10&24+10&24+10&24+10&24+10&24+10&24+10&24+10&24+10&24+10&24+10\\
36&24+12&26+10&24+12&24+12&24+12&24+12&24+12&24+12&24+12&24+12&24+12&26+10&24+12\\
38&24+14&26+12&24+14&24+14&24+14&24+14&24+14&24+14&24+14&24+14&24+14&26+12&24+14\\
40&26+14&26+14&26+14&26+14&26+14&26+14&26+14&26+14&26+14&26+14&26+14&26+14&26+14\\
42&28+14&28+14&28+14&28+14&28+14&28+14&28+14&28+14&28+14&28+14&28+14&28+14&28+14\\

\hline
                                                                    
\end{tabular}

}

\end{table}

The most important idea in the present work is that the choice of the irrep possessing the highest eigenvalue of $C_2^{SU(3)}$ is justified up to the middle of the valence proton (neutron) shell, while in the upper half of the shell the highest weight irrep (hw irrep) \cite{code}, which is the irrep most probable to appear, should be used. This observation has been already made 
within the proxy-SU(3) scheme \cite{proxy1,proxy2}, which is an alternative scheme of restoring the SU(3) symmetry of the nuclear shells beyond the sd shell, and has been attributed to the short range nature 
of the nucleon-nucleon interaction \cite{Ring,Casten}, which forces the spatial part of the wave function to be as symmetric as possible, while the spin-isospin part of the wave functions remains as antisymmetric as possible \cite{Ring,Casten}. We show here that the choice of the hw irreps beyond mid-shells in the pseudo-SU(3) scheme leads to 
a prediction for the prolate to oblate shape transition in heavy rare earths which is in agreement with existing experimental evidence \cite{Namenson,Alkhomashi,Wheldon,Podolyak,Linnemann} recently reviewed in \cite{proxy2}, while the choice of the highest eigenvalue of $C_2^{SU(3)}$ within the whole shell leads to a transition from prolate to oblate shapes in the middle of the shell, which is not seen experimentally. 
In particular, $^{192}_{76}$Os$_{116}$ \cite{Namenson} and $^{190}_{74}$W$_{116}$ \cite{Alkhomashi} have been suggested as lying at the prolate-oblate border, with $^{194}_{76}$Os$_{118}$ \cite{Wheldon} and $^{198}_{76}$Os$_{122}$ \cite{Podolyak} having an oblate character. Data on nuclei from Hf to Pt, discussed in Ref. \cite{Linnemann}, also suggests that the transition occurs between $^{192}_{76}$Os$_{116}$ and $^{194}_{78}$Pt$_{116}$. In other words, there is experimental evidence converging on a prolate to oblate transition occurring around $N=116$ for the W, Os, and Pt series of isotopes, which is in agreement with the findings of the present work, as we shall see below. 

\begin{table}[htb]

\caption{Highest weight irreducible representations (irreps), labeled by hw, and irreps possessing the highest eigenvalue of the second order Casimir operator of SU(3) (see Eq. (\ref{C2})), labeled by C,   
occurring in the decomposition of U(10) and U(15) for $M$ particles, as obtained through the code of Ref. \cite{code}. Oblate irreps are shown in boldface. A more extended version of the table has been given in Ref. \cite{proxy2}.
}

\bigskip
\begin{tabular}{ r c c c c c c c   }

\hline
  M     & 2 & 4 & 6 & 8 & 10 & 12 & 14 \\
\hline 
U(10) hw&(6,0)&(8,2)&(12,0)&(10,4)&(10,4)&(12,0)&(6,6)\\
U(10) C &(6,0)&(8,2)&(12,0)&(10,4)&(10,4)&{\bf (4,10)}&{\bf(0,12)}\\
U(15) hw&(8,0)&(12,2)&(18,0)&(18,4)&(20,4)&(24,0)&(20,6)\\
U(15) C &(8,0)&(12,2)&(18,0)&(18,4)&(20,4)&(24,0)&(20,6)\\
\hline

\end{tabular}

\begin{tabular}{ r c c c c c c c   }

\hline
  M     &   16 & 18 & 20 & 22 & 24 & 26 & 28  \\
\hline 
U(10) hw & {\bf (2,8)}&{\bf(0,6)}&(0,0)&   &   &  &   \\
U(10) C & {\bf(2,8)}&{\bf(0,6)}&(0,0)&  &   &  &   \\
U(15) hw & (18,8)&(18,6)&(20,0)&(12,8)&{\bf (6,12)}&{\bf(2,12)}&{\bf(0,8)}\\
U(15) C &{\bf (6,20)}&{\bf(0,24)}&{\bf(4,20)}&{\bf(4,18)}&{\bf(0,18)}&{\bf(2,12)}&{\bf(0,8)}\\
\hline

\end{tabular}

\end{table}

\begin{table}[htb]

\caption{Total irreps corresponding to rare earth nuclei obtained when the highest weight irreps are used for both the valence protons and the valence neutrons. 
The irreps are taken from Table 2, as explained in the text through two examples. Oblate irreps are shown in boldface.
}

\bigskip

\rotatebox{90}{

\begin{tabular}{ r c c c c c c c c c c c c c }

\hline
N & Xe & Ba & Ce & Nd & Sm & Gd & Dy & Er & Yb & Hf & W & Os & Pt \\
\hline
 84& 16,2& 20,0& 20,0& 20,0& 18,4& 18,4& 18,4& 18,4& 20,0& 14,6& 10,8& 10,8& 10,8\\
 86& 20,4& 24,2& 24,2& 24,2& 22,6& 22,6& 22,6& 22,6& 24,2& 18,8&14,10&14,10&14,10\\
 88& 26,2& 30,0& 30,0& 30,0& 28,4& 28,4& 28,4& 28,4& 30,0& 24,6& 20,8& 20,8& 20,8\\
 90& 26,6& 30,0& 30,0& 30,0& 28,4& 28,4& 28,4& 28,4& 30,0&24,10&20,12&20,12&20,12\\
 92& 26,6& 30,0& 30,0& 30,0& 28,4& 28,4& 28,4& 28,4& 30,0&24,10&20,12&20,12&22,12\\
 94& 26,6& 30,4& 30,4& 30,4& 28,8& 28,8& 28,8& 28,8& 30,4&26,10&20,12&20,12&22,12\\
 96& 28,6& 30,4& 30,4& 30,4& 28,8& 28,8& 28,8& 28,8& 30,4&26,10&22,12&20,12&22,12\\
 98& 28,6& 32,4& 32,4& 32,4& 30,8& 30,8& 30,8& 30,8& 32,4& 30,6&22,12&22,12&22,12\\
100& 32,2& 36,0& 36,0& 36,0& 34,4& 34,4& 34,4& 34,4& 36,0& 30,6& 26,8& 26,8& 26,8\\
102& 32,2& 36,0& 36,0& 36,0& 34,4& 34,4& 34,4& 34,4& 32,6&26,12& 26,8& 26,8& 26,8\\
104& 28,8& 32,6& 32,6& 32,6&30,10&30,10&30,10&30,10& 32,6&24,14&22,14&22,14&22,14\\
106&26,10& 30,8& 30,8& 30,8&28,12&28,12&28,12&28,12& 30,8&24,14&20,16&22,14&20,16\\
108&26,10& 30,8& 30,8& 30,8&28,12&28,12&28,12&28,12& 30,6&24,14&20,16&20,16&20,16\\
110& 26,8& 30,6& 30,6& 30,6&28,10&28,10&28,10&28,10& 30,6&24,12&20,14&20,14&20,14\\
112& 28,2& 32,0& 32,0& 32,0& 30,4& 30,4& 30,4& 30,4& 32,0& 26,6& 22,8& 22,8& 22,8\\
114&20,10& 24,8& 24,8& 24,8&22,12&22,12&22,12&22,12& 24,8&18,14&{\bf 14,16}&{\bf 14,16}&{\bf 14,16}\\
116&14,14&18,12&18,12&18,12&16,16&16,16&16,16&16,16&18,12&{\bf 12,18}&{\bf 8,20}&{\bf 8,20}&{\bf 8,20}\\
118&14,14&14,12&18,12&18,12&16,16&16,16&16,16&16,16&18,12&{\bf 12,18}&{\bf 8,20}&{\bf 4,20}&{\bf 8,20}\\
120&14,14&14,12&18,12&18,12&16,16&16,16&16,16&16,16&18,12&{\bf 12,18}& {\bf 8,20}&{\bf  4,20}&{\bf 8,20}\\
122&{\bf 10,14}&14,12&14,12&14,12&{\bf 12,16}&{\bf 12,16}&{\bf 12,16}&{\bf 12,16}&14,12& {\bf 8,18}& {\bf 4,20}&{\bf 4,20}&{\bf 4,20}\\
124&{\bf 8,10}& 12,8& 12,8& 12,8&{\bf 10,12}&{\bf 10,12}&{\bf 10,12}& {\bf 10,12}& 12,8& {\bf 6,14}&{\bf 2,16}& {\bf 2,16}& {\bf 2,16}\\

\hline

\end{tabular}

}

\end{table}

\begin{table}[htb]

\caption{Total irreps corresponding to rare earth nuclei obtained when the irrep having the highest eigenvalue  of the second order Casimir operator of SU(3)
is used for both the valence protons and the valence neutrons. The irreps are taken from Table 2, as explained in the text through two examples. Oblate irreps are shown in boldface.
}

\bigskip

 \rotatebox{90}{

\begin{tabular}{ r c c c c c c c c c c c c c }

\hline
N & Xe & Ba & Ce & Nd & Sm & Gd & Dy & Er & Yb & Hf & W & Os & Pt \\
\hline
 
 84& 16,2& 20,0& 20,0& 20,0& 18,4& 18,4& 18,4& 18,4&12,10& 8,12& 10,8& 10,8& 10,8\\
 86& 20,4& 24,2& 24,2& 24,2& 22,6& 22,6& 22,6& 22,6&16,12&12,14&14,10&14,10&14,10\\
 88& 26,2& 30,0& 30,0& 30,0& 28,4& 28,4& 28,4& 28,4&22,10&18,12& 20,8& 20,8& 20,8\\
 90& 26,6& 30,0& 30,0& 30,0& 28,4& 28,4& 28,4& 28,4&22,10&18,16&20,12&20,12&20,12\\
 92& 26,6& 30,0& 30,0& 30,0& 28,4& 28,4& 28,4& 28,4&22,10&18,16&20,12&20,12&22,12\\
 94& 26,6& 30,4& 30,4& 30,4& 28,8& 28,8& 28,8& 28,8&22,14&20,16&20,12&20,12&22,12\\
 96& 28,6& 30,4& 30,4& 30,4& 28,8& 28,8& 28,8& 28,8&22,14&20,16&22,12&20,12&22,12\\
 98& 28,6& 32,4& 32,4& 32,4& 30,8& 30,8& 30,8& 30,8&24,14&24,12&22,12&22,12&22,12\\
100& 32,2& 36,0& 36,0& 36,0& 34,4& 34,4& 34,4& 34,4&28,10&24,12& 26,8& 26,8& 26,8\\
102& 32,2& 36,0& 36,0& 36,0& 34,4& 34,4& 34,4& 34,4&24,16&20,18& 26,8& 26,8& 26,8\\
104& 28,8& 32,6& 32,6& 32,6&30,10&30,10&30,10&30,10&24,16&{\bf 6,32}&22,14&22,14&22,14\\ 
106&{\bf 14,22}&{\bf 18,20}&{\bf 18,20}&{\bf 18,20}&{\bf 16,24}&{\bf 16,24}&{\bf 16,24}&{\bf 16,24}&{\bf 10,30}& {\bf 6,32}& {\bf 8,28} &{\bf 22,14}& {\bf 8,28}\\
108&{\bf 14,22} &{\bf 18,20} &{\bf 18,20}&{\bf 18,20}&{\bf 16,24}&{\bf 16,24}&{\bf 16,24}&{\bf 16,24}&{\bf 4,34}&{\bf 6,32}&{\bf 8,28}&{\bf 8,28}&{\bf 8,28}\\
110&{\bf 8,26}&{\bf 12,24}&{\bf 12,24}&{\bf 12,24}&{\bf 10,28}&{\bf 10,28}&{\bf 10,28}&{\bf 10,28}&{\bf  4,34}&{\bf 0,36}&{\bf 2,32}&{\bf 2,32}&{\bf 2,32}\\
112&{\bf 12,22}&{\bf 16,20}&{\bf 16,20}&{\bf 16,20}&{\bf 14,24}&{\bf 14,24}&{\bf 14,24}&{\bf 14,24}&{\bf  8,30}&{\bf 4,32}&{\bf 6,28}&{\bf 6,28}&{\bf  6,28}\\
114&{\bf 12,20}&{\bf 16,18}&{\bf 16,18}&{\bf 16,18}&{\bf 14,22}&{\bf 14,22}&{\bf 14,22}&{\bf 14,22}&{\bf  8,28}& {\bf 4,30}& {\bf 6,26}& {\bf 6,26}&{\bf 6,26}\\
116&{\bf 8,20}&{\bf 12,18}&{\bf 12,18}&{\bf 12,18}&{\bf 10,22}&{\bf 10,22}&{\bf 10,22}&{\bf 10,22}&{\bf 4,28}&{\bf 0,30}&{\bf 2,26}&{\bf 2,26}&{\bf 2,26}\\
118&{\bf 8,20}&14,12&{\bf 12,18}&{\bf 12,18}&{\bf 10,22}&{\bf 10,22}&{\bf 10,22}&{\bf 10,22}&{\bf 4,28}& {\bf 0,30}&{\bf 2,26}&{\bf 4,20}&{\bf 2,26}\\
120&{\bf 8,20}&14,12&{\bf 12,18}&{\bf 12,18}&{\bf 10,22}&{\bf 10,22}&{\bf 10,22}&{\bf 10,22}&{\bf 4,28}& {\bf 0,30}&{\bf 2,26}&{\bf 4,20}&{\bf 2,26}\\
122&{\bf 10,14}&14,12&14,12&14,12&{\bf 12,16}&{\bf 12,16}&{\bf 12,16}&{\bf 12,16}&{\bf 6,22}&{\bf 2,24}& {\bf 4,20}&{\bf 4,20}&{\bf 4,20}\\
124&{\bf 8,10}& 12,8& 12,8& 12,8&{\bf 10,12}&{\bf 10,12}&{\bf 10,12}&{\bf 10,12}&{\bf 4,18}&{\bf 0,20}& {\bf 2,16}&{\bf 2,16}&{\bf 2,16}\\

\hline

\end{tabular}

}

\end{table} 

The choice of the hw irreps instead of the irreps with the highest eigenvalue of $C_2^{SU(3)}$ offers 
an interesting by-product, since it turns out that in most nuclei the resulting total irrep for protons 
and neutrons is prolate, while oblate nuclei occur only near the end of the proton and neutron shells.
This result suggests an answer to the old standing question of the prolate over oblate dominance in even-even nuclei \cite{Hamamoto}, as also seen within the proxy-SU(3) scheme \cite{proxy2,proxy3}. 

We consider in detail the rare earths with 50-82 protons and 82-126 neutrons. The distribution of valence protons and neutrons into orbitals of normal and abnormal parity is shown in Table 1. The normal parity protons belong to a pseudo-pf shell possessing a U(10) overall symmetry, while the normal parity neutrons belong to a pseudo-sdg shell having an overall U(15) symmetry \cite{DW1,pseudo1,pseudo2}. One can read the appropriate irrep for protons and for neutrons from Table 2, obtained through the use of the code of Ref. \cite{code}. The resulting overall irrep for each nucleus  obtained when the highest weight irrep is used 
for both the valence protons and the valence neutrons is shown in Table 3, while in Table 4 
the resulting overall irrep for each nucleus  obtained when the irrep with the highest $C_2^{SU(3)}$ eigenvalue is used for both the valence protons and the valence neutrons is shown. 

Two examples are given for illuminating purposes. 

a) For $^{154}_{62}$Sm$_{92}$ one sees in \cite{Gogny} that the expected deformation is $\beta=0.342$, 
which is very close to the experimental value of 0.340 reported in \cite{Pritychenko}. 
The deformation parameter $\epsilon$ of the Nilsson model is related to $\beta$ through the equation 
$\epsilon=0.946 \beta$ \cite{Nilsson2}. Looking at the standard proton Nilsson diagrams for $\epsilon=0.32$
\cite{Lederer} we see that the 12 valence protons of $^{154}$Sm are occupying 4 orbitals of normal parity 
and 2 orbitals of abnormal parity. Similarly, looking at the standard neutron Nilsson diagrams for $\epsilon=0.32$
\cite{Lederer} we see that the 10 valence neutrons of $^{154}$Sm are occupying 3 orbitals of normal parity 
and 2 orbitals of abnormal parity. These values are reported in Table 1, taking into account that each orbital accommodates two particles. Now from Table 2 one sees that 
8 protons (the ones with normal parity) in the U(10) shell correspond to the irrep (10,4), while 
6 neutrons (the ones with normal parity) in the U(15) shell correspond to the irrep (18,0). Therefore the total irrep for $^{154}$Sm, reported in Table 3, is (28,4). Notice that since both the valence protons and neutrons of normal parity lie within the first half of their corresponding shell, the choice of the hw irreps vs. the irreps with highest eigenvalue of $C_2^{SU(3)}$ makes no difference, thus the same total irrep appears also in Table 4.   

b)For $^{180}_{72}$Hf$_{108}$ one sees in \cite{Gogny} that the expected deformation is $\beta=0.299$, 
which is close to the experimental value of 0.273 reported in \cite{Pritychenko}. 
 Looking at the standard proton Nilsson diagrams for $\epsilon=0.28$
\cite{Lederer} we see that the 22 valence protons of $^{180}$Hf are occupying 7 orbitals of normal parity 
and 4 orbitals of abnormal parity. Similarly, looking at the standard neutron Nilsson diagrams for $\epsilon=0.28$
\cite{Lederer} we see that the 26 valence neutrons of $^{180}$Hf are occupying 8 orbitals of normal parity 
and 5 orbitals of abnormal parity. These values are reported in Table 1. Now from Table 2 one sees that 
14 protons (the ones with normal parity) in the U(10) shell correspond to the hw irrep (6,6),
but they belong to the (0,12) irrep if the highest eigenvalue of $C_2^{SU(3)}$ is considered. Furthermore
16 neutrons (the ones with normal parity) in the U(15) shell correspond to the hw irrep (18,8),
but they belong to the (6,20) irrep if the highest eigenvalue of $C_2^{SU(3)}$ is considered.
Therefore the total irrep for $^{180}$Hf is (24,14) if the hw irreps are taken into account and is reported in Table 3. However, if the irreps with the highest eigenvalues of $C_2^{SU(3)}$ are considered,
the total irrep for the same nucleus is (6,32), reported in Table 4. Notice that since both the valence protons and neutrons of normal parity lie within the second half of their corresponding shell, the choice of the hw irreps vs. the irreps with highest eigenvalue of $C_2^{SU(3)}$ makes a big difference, resulting in a clearly prolate irrep in the first case and in a clearly oblate irrep in the second case.  

From Table 4 it is already clear that in the case in which the irreps with the highest $C_2^{SU(3)}$ eigenvalues are chosen, a transition from prolate irreps (having $\lambda > \mu$) to oblate irreps
(having $\lambda<\mu$) occurs exactly in the middle of the shell, while from Table 3 it is seen that in the case of choosing the hw irreps beyond mid-shell, the prolate to oblate transition is observed at substantially higher neutron numbers.
Furthermore, in the highest $C_2^{SU(3)}$ eigenvalue case all series of isotopes from Xe to Pt exhibit 
oblate irreps, while in the hw irreps case only the Hf, W, Os and Pt series of isotopes do so.  
This effect can be visualized in Fig. 1, in which the predictions for the $\gamma$ collective deformation parameter 
are plotted as a function of the neutron number $N$ for the Xe to Pt chains of isotopes under study.
The values of $\gamma$ are obtained from \cite{Castanos,Park}
\begin{equation}\label{g1}
\gamma = \arctan \left( {\sqrt{3} (\mu+1) \over 2\lambda+\mu+3}  \right),
\end{equation}
which is parameter independent. Indeed in Fig. 1 we see that in the case of the hw irreps the $\gamma$ values are suddenly starting to jump up at $N=114$, while the jump is completed at $N=116$, in agreement 
with the experimental evidence \cite{Namenson,Alkhomashi,Wheldon,Podolyak,Linnemann} mentioned above.
After $N=116$ the $\gamma$ values are clearly stabilized, as seen in the hw irrep panel of Fig. 1. 
Furthermore, only the Hf, W, Os, Pt series of isotopes enter the oblate region ($\gamma>30^{\rm{o}}$), in agreement 
with the experimental evidence. In contrast, in the highest $C_2^{SU(3)}$ eigenvalue case, all series of isotopes from Xe to Pt jump into the oblate region at the neutron midshell.


\begin{figure}[htb]

{\includegraphics[width=75mm]{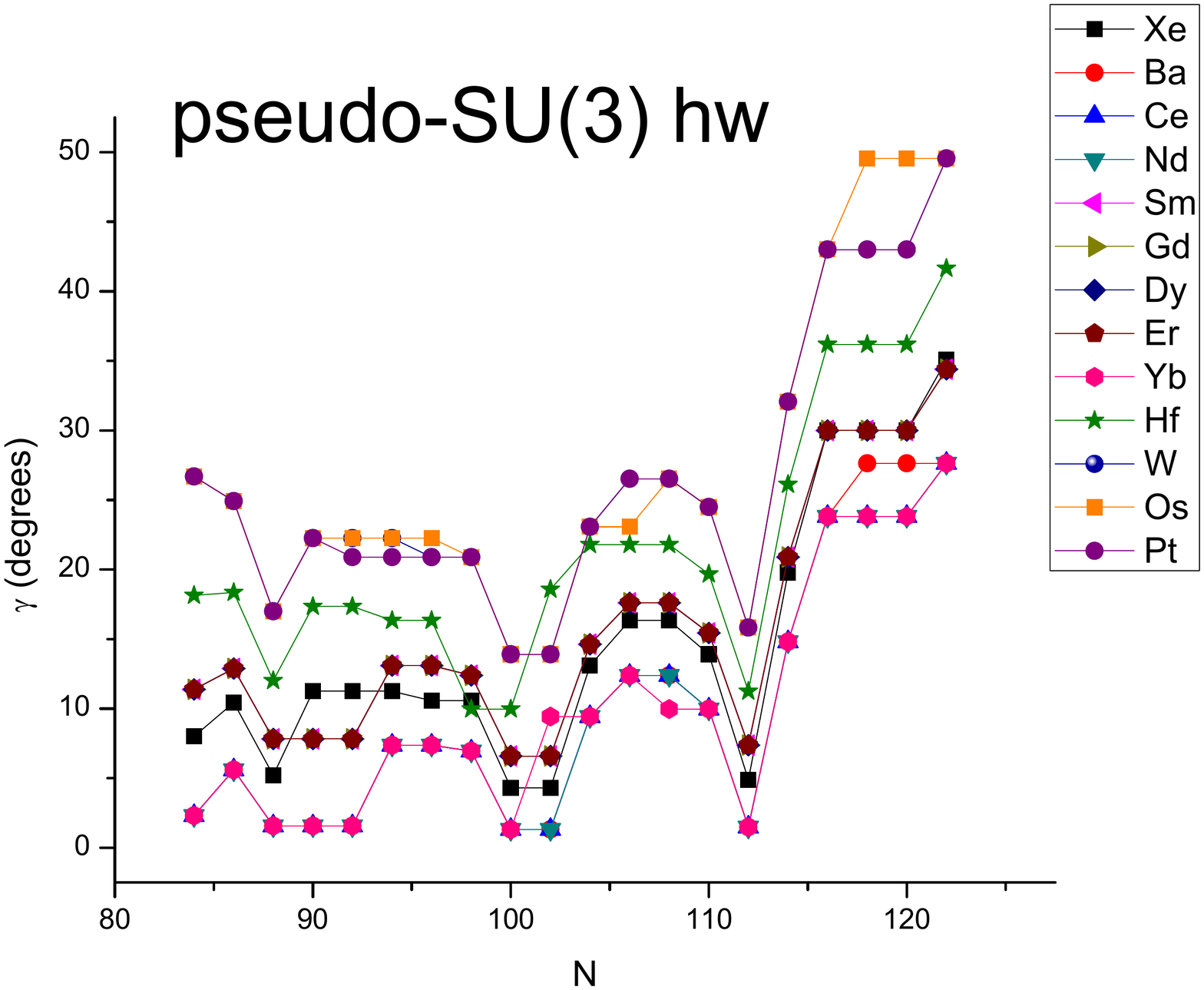}\hspace{5mm}
\includegraphics[width=75mm]{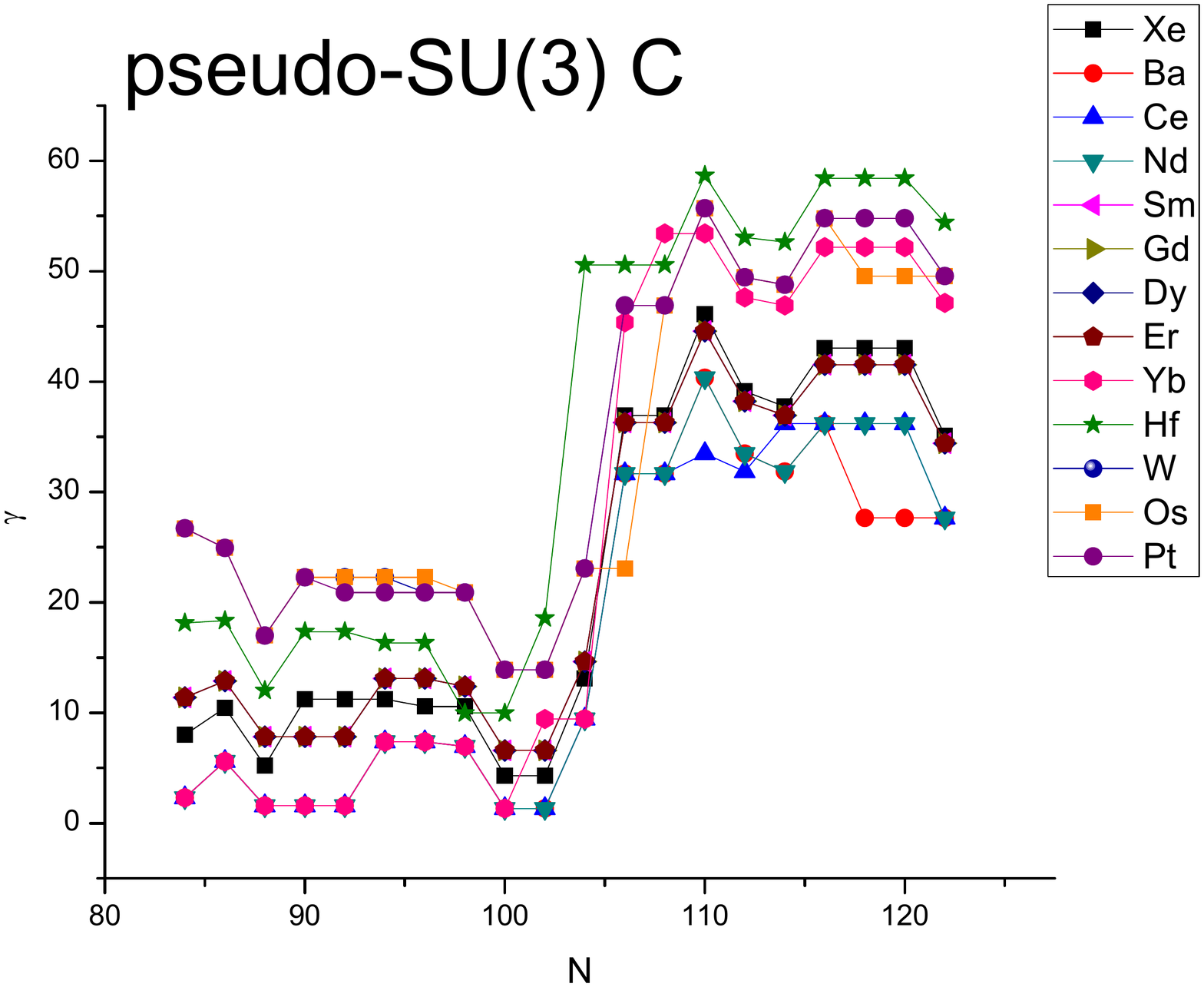}}

\caption{Pseudo-SU(3) predictions for the collective deformation parameter $\gamma$, as obtained from the irreps of Tables 3 (labeled as pseudo-SU(3) hw) and 4 (labeled as pseudo-SU(3) C) using Eq. (\ref{g1}). See the text for further discussion. 
} 

\end{figure}

In Fig. 1 it is also seen that in the case of the hw irreps the prolate nuclei (with $\gamma<30^{\rm{o}}$)
greatly outnumber the oblate nuclei (with $\gamma>30^{\rm{o}}$), thus offering an answer to the old-standing question \cite{Hamamoto} of prolate over oblate dominance in the ground states of even-even nuclei. 

In conclusion, it is seen that within the framework of the pseudo-SU(3) scheme one can predict the prolate to oblate shape transition occurring in heavy rare earths simply by taking into account that beyond the middle of a harmonic oscillator shell possessing an SU(3) subalgebra the highest weight irreducible representation (hw irrep) has to be used instead of the irrep having the highest eigenvalue of the second order Casimir operator of SU(3), $C_2^{SU(3)}$. This choice is a consequence of the short range of the nucleon-nucleon interaction, as discussed in detail in \cite{GC40}, in relation to the proxy-SU(3) scheme. It is interesting that two different approximation schemes restoring SU(3) in the nuclear shells beyond the sd shell yield similar results for the prolate to oblate shape transition, free of any free parameters, by simply taking into account the consequences of the short range nature of the nucleon-nucleon interaction, which requires the spatial part of the wave function to be as symmetric as possible, as discussed in detail in \cite{GC40}. The prolate over oblate dominance in the ground states of even-even-nuclei also comes out
as a by-product of the choice of the hw irreps instead of the irreps with the highest eigenvalue 
of $C_2^{SU(3)}$.  

It should be pointed out that the necessity to use the hw irreps beyond midshell is a general feature of strongly interacting fermionic systems occupying finite shells, which is not limited within the realm of the pseudo-SU(3) and proxy-SU(3) schemes. 

It should be emphasized that the numerous applications of the pseudo-SU(3) scheme in many different 
nuclear physics problems \cite{DW1,DW2,Vargas1,Vargas2} have so far been limited in the first half of the relevant harmonic oscillator shells, 
thus they are not influenced by the present findings, which just pave the way for the application 
of the pseudo-SU(3) scheme beyond the middle of the relevant harmonic oscillator shells.

Several further steps can be considered. 

a) For each chain of isotopes the results for $\gamma$ obtained by pseudo-SU(3) can be compared to predictions by the 
Gogny D1S interaction \cite{Gogny}. In addition, they can be compared to existing experimental values, as in Ref.  \cite{proxy2}.  

b) The collective deformation parameter $\beta$ can also be calculated from the irreps of Table 3, 
again in a parameter free way \cite{proxy2,Castanos,Park}. A scaling factor related to the fractions of the shells used will enter in this case \cite{proxy2}.

c) For each chain of isotopes the results for $\beta$ obtained by pseudo-SU(3) can be compared to predictions by the 
Gogny D1S interaction \cite{Gogny}, as well as to predictions by Relativistic Mean Field (RMF) theory
\cite{Lalazissis}. In addition,
they can be compared to existing experimental values \cite{Pritychenko}.

Support by the BNSF under contract No. KP-06-N28/6 is gratefully acknowledged by N.M.

\end{document}